\documentclass[onecolumn,fleqn,usenatbib]{mnras}

\usepackage[T1]{fontenc}
\usepackage{ae,aecompl}
\usepackage{hyperref}

\usepackage{graphicx}    % Including figure files
\usepackage{amsmath}    % Advanced maths commands
\usepackage{amssymb}    % Extra maths symbols

\usepackage{orcidlink}

\newcommand{\source}{G\ensuremath{118.4+37.0}}

\title[\textit{Fermi}-LAT detection of G$118.4+37.0$]{\textit{Fermi}-LAT detection of G$118.4+37.0$: a supernova remnant in the Galactic halo seen around the Calvera pulsar}

\author[Araya]{
M.~Araya\,\orcidlink{0000-0002-0595-9267}$^{1}$\thanks{E-mail: miguel.araya@ucr.ac.cr},%orcid 0000-0002-0595-9267
\\
$^{1}$Escuela de F\'isica, Universidad de Costa Rica, Montes de Oca, San Jos\'e, Costa Rica, 11501-2060}

\date{Accepted ; Received ; in original form }

\pubyear{2022}

\begin{document}
\label{firstpage}
\pagerange{\pageref{firstpage}--\pageref{lastpage}}
\maketitle

\begin{abstract}
The discovery of a non-thermal radio ring of low surface brightness about one degree in diameter has been recently reported around the location in the sky of the Calvera pulsar, at a high Galactic latitude. The radio properties point to it likely being a new supernova remnant (SNR), \source. We report an analysis of almost 14 years of observations of this region by the $\gamma-$ray Large Area Telescope onboard the \textit{Fermi} satellite. We detect extended GeV emission consistent with the size and location of the radio source, which confirms the presence of relativistic particles. The spectrum of the high-energy emission is fully compatible with an origin in the same relativistic particles producing the radio emission. These features and its similarities to other isolated SNRs establish this source as the remnant of a supernova. A simple model of the non-thermal emission from radio to GeV energies resulting from leptonic emission from electrons produced by the SNR is presented. \source{} and other similar isolated remnants could be part of a radio-dim SNR population evolving in low density environments showing hard GeV emission of leptonic origin. Future deeper surveys in radio and $\gamma$-rays could discover new members of the group.
\end{abstract}


\section{Introduction}\label{sec:intro}
A supernova remnant (SNR) can result from the collapse of the core and subsequent explosion of a massive star as well as from the thermonuclear destruction of an accreting white dwarf. The explosion distributes heavy elements throughout the Galaxy, and its blast wave expands at speeds of thousands of kilometers per second, heating up the surrounding interstellar medium (ISM) to very high temperatures and accelerating cosmic rays to very high energies. The study of SNRs also provides constraints on the stellar and ISM evolution in the Galaxy.

Approximately 300 SNRs have been found in our Galaxy \citep[e.g.,][]{2019JApA...40...36G}, which is about twice the amount of SNRs that were known 30 years ago. However this number is significantly below the expected number ($\gtrsim 1000$) of SNRs based on OB star counts, pulsar birth rates, Fe abundances and the supernova rate in Local Group galaxies \citep{1991ApJ...378...93L,1994ApJS...92..487T}. The discrepancy could be due in part to a lack of sensitivity in the observations and thus it is important to carry out more detailed studies that can contribute to understanding stellar evolution. In fact, new supernova remnants have been found in recent surveys. For example a study using The HI, OH, Recombination line survey of the Milky Way (THOR), in combination with lower-resolution VLA 1.4 GHz Galactic Plane Survey (VGPS) continuum data, MIR data from the Spitzer GLIMPSE, Spitzer MIPSGAL, and WISE surveys has recently reported 76 new Galactic SNR candidates \citep{2017A&A...605A..58A}. Other studies such as the GaLactic and Extragalactic All-sky Murchison Widefield Array (GLEAM) survey \citep{2019PASA...36...48H}, and the Molonglo Galactic Plane Survey \citep{2014PASA...31...42G} have also discovered new SNRs.

$\gamma-$ray surveys can also be useful to detect previously unknown SNRs since the associated cosmic rays produce high and very-high energy emission (from GeV to TeV energies). Relativistic electrons in SNRs can produce $\gamma-$rays if they interact with soft photons through the process of inverse Compton, or as a result of bremsstrahlung interactions with gas (this is known as leptonic emission). Relativistic protons and nuclei, on the other hand, can produce GeV to TeV photons through inelastic collisions with ambient gas (the hadronic emission).

The Large Area Telescope (LAT) onboard the \textit{Fermi} satellite \citep{2009ApJ...697.1071A} continuously scans the sky in the GeV regime and has revealed a variety of sources and structures in the sky, including SNRs \citep[see, for example,][]{2016ApJS..224....8A}. Besides well known SNRs the LAT has shown radio dim SNRs can also shine in GeV energies. Examples include the SNRs G$150.3+4.5$, G$279.0+1.1$ and G$323.7-1.0$ \citep{2020A&A...643A..28D,2020MNRAS.492.5980A,2017ApJ...843...12A}, which show a hard GeV spectrum. Other extended sources with similar GeV spectra, although of unknown nature, include G$350.6-4.7$ and 2HWC~J$2006+341$ \citep{2018MNRAS.474..102A,2020ApJ...903L..14A}. It is likely that these unidentified sources will be associated to new SNRs or pulsar wind nebulae (PWN) in the future. More recently, \cite{2022MNRAS.510.2920A} reported the discovery of a radio dim SNR, G$17.8 + 16.7$, located at a high Galactic latitude. The source, approximately circular in shape with a diameter of $\sim 0.8^\circ$, shows a radio spectral index $\alpha= -0.75 \pm 0.15$ and a GeV spectrum characterized by a simple power law of the form $\frac{dN}{dE}\propto E^{-\Gamma}$ with $\Gamma \sim 1.8$.

\cite{2022arXiv220714141A} have recently reported the discovery of a dim radio ring at a high Galactic latitude ($b=37^\circ$) with data from the LOFAR Two-meter Sky Survey \citep[LoTSS,][]{2017A&A...598A.104S}. The outer and inner radii of the radio ring are $29'$ and $14'$. The centre of the roughly circular structure is located $4.87'$ from the position in the sky of the neutron star 1RXS J$141256.0+792204$ (PSR J$1412+7922$). The radio spetral index calculated from their observations is $\alpha = -0.71\pm 0.09$ ($S_\nu \propto \nu^\alpha$), within the usual range for synchrotron emission from relativistic electrons in an SNR. The integrated flux density of the ring is low, $0.22\pm0.04$ Jy at a frequency of 1420 MHz. A small region of extended H$\alpha$ emission internal to the ring was also detected \citep{2022arXiv220714141A}. A scenario where the radio structure is the remnant of the supernova that formed 1RXS J$141256.0+792204$ is favoured by the authors that reported its discovery. Regardless of its relation to this neutron star, the SNR nature of the source is also supported by the discovery of diffuse soft X-ray emission 13 arcmin west of the pulsar position whose spectrum is compatible with that of an evolved SNR \citep{2011MNRAS.410.2428Z}.

The neutron star 1RXS J$141256.0+792204$ seen near the centre of the radio ring was nicknamed ``Calvera'' due to its similarities with the seven nearby X-ray dim isolated neutron stars (XDINSs) known as the ``Magnificent Seven'' \citep{2007Ap&SS.308..181H}. The star actually displays properties that are unlike those of XDINSs and are more similar to those of young radio pulsars, besides being located unusually far from the Galactic disk. Calvera is detected only through its thermal X-ray emission, it has a spin-down luminosity $\dot{E} = 6.1\times 10^{35}$~erg~s$^{-1}$, a characteristic age $\tau_c = 2.9\times 10^5$~yr and a spin period of 59~ms \citep{2013ApJ...778..120H,2019ApJ...877...69B}. \cite{2021ApJ...922..253M} set a distance of 3.3~kpc (3.1--3.8~kpc) to the pulsar from their best fit to the thermal X-ray emission, and found no pulsed $\gamma-$rays from the star. This distance estimate and upper limits on its high-energy flux make Calvera underluminous in $\gamma-$rays \citep[$<7\times10^{32}\times d^2_{\mbox{\tiny 3 kpc}}$~erg~s$^{-1}$,][]{2013ApJ...778..120H} compared to other pulsars with similar spin-down power, although less so than originally thought. \cite{2021ApJ...922..253M} conclude that Calvera could have been born in the Galactic halo from the explosion of a runaway massive star or the accretion-induced collapse of a white dwarf.

In this work we report the detection of GeV $\gamma-$rays consistent with the extension and location of the radio structure, confirming the presence of relativistic particles in the source. The properties of the high-energy emission, such as its extent and spectrum, as well as its similarities to other isolated remnants such G$17.8 + 16.7$, confirm its nature as an SNR, \source. In section \ref{sec:data} we describe the analysis of LAT data revealing the source at high energies while in section \ref{discussion} we discuss the origin of the $\gamma-$rays, which are likely produced by the same particles responsible for the non-thermal radio emission observed.

\section{LAT data}\label{sec:data}
The \emph{Fermi}-LAT is a converter/tracker telescope detecting gamma rays in the energy range between 20~MeV and $\ga$1~TeV \citep{2009ApJ...697.1071A}. We combined observations from October 2008 to July 2022 in the energy range 100~MeV--500~GeV. The data gathered during the first two months of the mission (prior to the Mission Elapsed Time 246823875), which suffered from high levels of background contamination for energies above $\sim 30$ GeV\footnote{See \url{https://fermi.gsfc.nasa.gov/ssc/data/analysis/LAT\_caveats.html}}, were not used. The {\tt Pass 8} dataset was analyzed with the software {\tt fermitools} version~2.2.0 by means of the {\tt fermipy} package version~1.1.6. We selected good quality front and back-converted events in the {\tt SOURCE} class ({\tt evclass=128}, {\tt evtype=3}) having zenith angles lower than $90\degr$ to avoid contamination from $\gamma-$rays from Earth's limb.

The response functions corresponding to the dataset used for the analysis are {\tt P8R3\_SOURCE\_V3}. We binned the data with a spatial scale of $0.05\degr$ per pixel and ten bins in energy for exposure calculations. To model the background sources we used the latest incremental version (4FGL-DR3) of the fourth catalog of LAT sources which is based on 12 years of survey data in the 50 MeV--1 TeV energy range \citep{2020ApJS..247...33A,2022ApJS..260...53A}. A LAT source labeled 4FGL~J1409.8+7921 recently reported in the 4FGL-DR3 and found within the extent of \source{} was not included in the model of the region since we carried out a more detailed study of the GeV emission. In the LAT catalog this source is described as a dim non-variable point source with a spectrum consistent with a simple power law function and a spectral index of $1.82\pm0.22_{\mbox{\tiny stat}}$.

The diffuse Galactic emission and the isotropic emission (including the residual cosmic-ray background) were modeled with the files {\tt gll\_iem\_v07.fits} and {\tt iso\_P8R3\_SOURCE\_V3\_v1.txt}, respectively, provided by the LAT team\footnote{See \url{https://fermi.gsfc.nasa.gov/ssc/data/access/lat/BackgroundModels.html}}. Also as recommended by the LAT team, the energy dispersion correction was applied to all sources except for the isotropic component\footnote{See \url{https://fermi.gsfc.nasa.gov/ssc/data/analysis/documentation/Pass8\_edisp\_usage.html}}. The emission from each source is convolved with the Instrument Response Functions using the {\tt fermitools} to predict the number of observed counts in a spatial and energy binning. The maximum likelihood technique \citep{1996ApJ...461..396M} was used to obtain the morphological and spectral parameters of the sources that maximize the probability for the model to explain the data. The detection significance of a source can be calculated using the test statistic (TS), defined as $-2\log(\mathcal{L}_0/\mathcal{L})$. Here $\mathcal{L}$ and $\mathcal{L}_0$ are the maximum likelihood functions for a model with the source and for the model without the additional source, which is known as the null hypothesis, respectively.

As a first step we carried out a morphological study of the emission using events having reconstructed energies above 1~GeV. This energy threshold was chosen to obtain sufficient statistics for the analysis while taking advantage of the improved spatial resolution of the LAT at higher energies. The 68\% containment angle of the PSF for front+back-converted events is $\sim0.8^\circ$ at 1~GeV and drops below $\sim0.15^\circ$ above 10~GeV, compared to $\sim5^\circ$ at 100~MeV. The region of interest (RoI) for the analysis included events reconstructed within $15\degr$ of the coordinates (J2000) RA = $212.8^\circ$, Dec = $79.4^\circ$, a location close to the centre of the radio ring. Cataloged sources found within $20\degr$ of this position were added to the model for the morphological analysis. We only allowed the spectral normalizations free to vary for the 4FGL sources found within $11\degr$ from the RoI centre. This way we made sure to fit the spectral normalization of the relatively bright active galactic nuclei S5~$1803+784$, seen about $10.6\degr$ from the centre of the region. For the sources found within $5\degr$ of the centre, all their spectral parameters, and not just the normalizations, were allowed to vary. In order to account for excess residuals in the region, possibly corresponding to un-modeled sources, we searched for hotspots having $\sqrt{\mbox{TS}}\geq 4$ with a minimum angular separation of $0.25\degr$ in the sky. Ten new point sources were found in the data, among them is one consistent with the location of the BL Lac NVSS~J$110105+860353$. Only two of the new point source (``PS'') candidates were found within $\sim6.9\degr$ of the centre of \source{} and are actually seen within the location of the radio ring, which we labeled PS~J$1409.5+7919$, seen close to the centre of \source{}, and PS~J$1403.0+7922$ in the western edge of the SNR. 

We calculated the Akaike Information Criterion \citep{1974ITAC...19..716A}, defined as AIC = $2k - 2\ln(\mathcal{L})$, where $k$ is the number of free parameters in the model, to compare the quality of the morphological models. The AIC takes into account the quality of the fit while incorporating a penalty related to the amount of free parameters in the fit. The geometrical models we compared are a 2D symmetric Gaussian, a uniform disk \citep[for their definition, see][]{2012ApJ...756....5L}, a uniform ring having the dimensions and location derived from the radio emission reported in the discovery of \source{} \citep{2022arXiv220714141A}, a single point source and two point sources, for which we optimized the locations with the dataset used. We fit the locations and extensions of the sources to obtain a likelihood profile scan in order to find the parameter set maximizing the likelihood. The spectrum of each source tested was assumed to be a simple power law, which is justified below. As indicated before, the model with two point sources included both PS~J$1409.5+7919$ and PS~J$1403.0+7922$ with resulting TS values of 22.1 and 18.6, respectively, while the model having one point source included only the most significant of the two, PS~J$1409.5+7919$.

The results can be seen in Table \ref{table:LAT}. The best morphological description of the emission among the tested models is given by the uniform disk. Models having a difference in AIC greater than 2 with respect to the lowest AIC value are considered significantly worst \citep{burnham2001}. The significance of the source extension can be calculated from TS$_{\mbox{\tiny ext}}$, which is equal to twice the difference between the $\log \mathcal{L}$ for an extended source model and that obtained by replacing the extended source with a point-like source whose position is optimized. Using the disk model we found the extended hypothesis to be preferred over a point source, with TS$_{\mbox{\tiny ext}} = 35.1$, confirming that the source is significantly extended. The best-fit radius found for the disk and its centre location are $0.53^{+0.03}_{-0.04}\degr$, RA$=212.42\pm0.06\degr$, Dec$=79.34\pm0.04\degr$ ($1\sigma$ statistical uncertainties quoted)\footnote{The corresponding Galactic coordinates are $l=118.48\pm 0.06\degr$, $b=37.10\pm0.04\degr$.}.

\begin{table*}
\caption{Results of the morphological analysis of \emph{Fermi}-LAT data.}
\label{table:LAT}
\begin{center}
\begin{tabular}{lccc}
\hline
\hline
Spatial model & Fitted size$^{a}$ ($\degr$) & $\Delta$AIC$^b$ \\
\hline
Disk & $0.53^{+0.03}_{-0.04}$ & 0\\
Gaussian & $0.41^{+0.06}_{-0.05}$ & 2.8 \\
Radio ring & Fixed & 9.0\\
1 point source & -- &  35.2\\
2 point sources & -- &  17.2\\
\hline
\end{tabular}\\
\textsuperscript{$a$}\footnotesize{Radius for the disk and 68\%-containment radius for the Gaussian and their $1\sigma$ statistical uncertainties.}\\
\textsuperscript{$b$}\footnotesize{$\Delta$AIC is equal to the value of AIC for each model minus the AIC value for the best-fit model.}\\
\end{center}
\end{table*}

\begin{figure}
    \centering
    \includegraphics[width=0.7\textwidth]{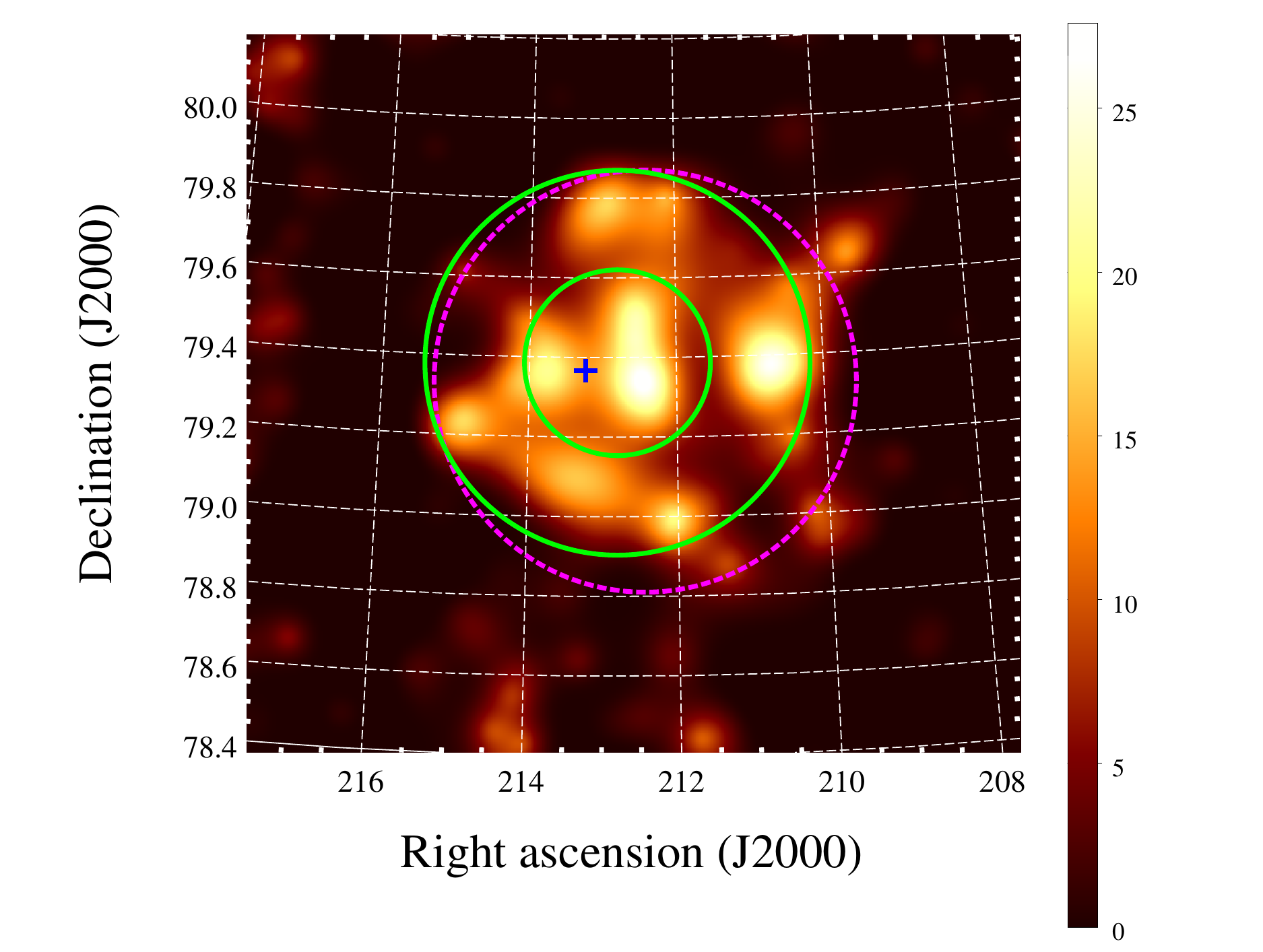}
    \caption{TS map (colour scale) obtained with LAT data in the energy range 1--500\,GeV. The image was generated by fitting the normalization of a point source placed at each pixel in the map and calculating its TS value. A simple power law spectrum of the form $\frac{dN}{dE}\propto E^{-1.8}$ was assumed for the point source. The solid circles represent the radio ring described by \protect\cite{2022arXiv220714141A} while the dashed circle represents the best-fit disk found in this work to model the $\gamma-$ray emission. The cross marks the position of the Calvera pulsar. Sky coordinates are shown in degrees.}
    \label{fig:tsmap}
\end{figure}

A TS map obtained from LAT data in the energy range 1--500\,GeV and without including any source in the model representing \source{} is shown in Fig. \ref{fig:tsmap}. This is therefore a map of the residual emission when the SNR is not accounted for in the model. Note this is a map for a ``point-source hypothesis''. It was obtained by fitting the spectral normalization of a point source (using a power-law hypothesis for its spectrum) in each pixel of the map and this way obtaining the corresponding TS values at each location, which the map shows.

Once the morphology of the source is properly modeled, we extracted events in five energy intervals equally spaced logarithmically in the range 100~MeV--500~GeV using the best-fit disk found and the baseline model of the region optimized before. In each interval we performed a binned likelihood analysis and modeled the spectrum of the source with a simple power law function using a fixed spectral index of 2, to find the spectral normalization. We concluded that the source is not significantly detected below an energy of 500~MeV (TS$<4$). We then explored the spectrum of the source in the 0.5--500~GeV energy range and obtained a flux upper limit in the 100--500~MeV range. For this analysis we used events within $20\degr$ of the source and included the cataloged sources located within $25\degr$ of the centre of \source{} in the model of the region. We started with the optimized background model for the sources in the region obtained in the morphological analysis above again leaving the normalizations of the sources located within 11$\degr$ of \source{} free as well as the spectral shape parameters of the sources found within 5$\degr$ of the same location.

We tested two different spectral functions to fit the differential spectrum of the $\gamma-$ray emission in the likelihood analysis: a simple power law of the form $\frac{dN}{dE}=N_0 \left( \frac{E}{E_0}\right)^{-\Gamma}$ and a log-parabola given by $\frac{dN}{dE}=N_0 \left( \frac{E}{E_0}\right)^{-(\alpha + \beta \,\mbox{log}(E/E_0))}$, where $E_0$ is a fixed scale factor. Since these are nested models with the more elaborate one containing an additional free parameter, the square root of the difference in TS values from the fits gives the significance of the log-parabola model over the simpler hypothesis \citep[the likelihood-ratio test, see][]{wilks1938}. We obtained an improvement at the $2\sigma$-level using the log-parabola, which is not significant. The spectral data can thus be described with a simple power law in the 0.5--500~GeV energy range. No significant residual emission is left when including the source in the model and the TS value obtained for \source{} in this energy range is 61.3.

We found the values for the spectral index and normalization $\Gamma=1.66\pm 0.10_{\mbox{\tiny stat}} \pm 0.03_{\mbox{\tiny sys}}$ and $N_0=(3.2\pm 0.5_{\mbox{\tiny stat}} \pm 0.2_{\mbox{\tiny sys}})\times 10^{-15}$~MeV$^{-1}$cm$^{-2}$s$^{-1}$, respectively (for $E_0=8409$~MeV). The systematic uncertainties were obtained considering two independent aspects of the analysis: the errors in the effective area of the LAT and the uncertainties in modeling the diffuse Galactic emission. We propagated the effective area uncertainties using a set of bracketing response functions as recommended by \cite{2012ApJS..203....4A}. In the fit using the standard effective area as well as in the alternative fits the pivot energy is used as the value of the scale parameter, $E_0=8409$~MeV. This value for the pivot energy was calculated with the propagated statistical uncertainty taking parameter correlations into account. For estimating the uncertainties in the model of the diffuse emission, we used the eight alternative model files developed originally by \cite{2016ApJS..224....8A}, scaled appropriately to account for differences in energy dispersion between {\tt Pass 7} and {\tt Pass 8} reprocessed data\footnote{See https://fermi.gsfc.nasa.gov/ssc/data/access/lat/Model\_details/Pass8\_rescaled\_model.html}. For each alternative Galactic diffuse emission model we fit the data to obtain the spectral index and normalization and estimated the systematic uncertainty as in \cite{2016ApJS..224....8A}. We obtained the total systematic uncertainty for each parameter adding both effects in quadrature. The systematic errors on the parameters from both effects are comparable.

%%%%Los errores totales son: 0.033 para el indice y 0.24 para el prefactor.

\begin{figure}
    \centering
    \includegraphics[width=0.7\textwidth]{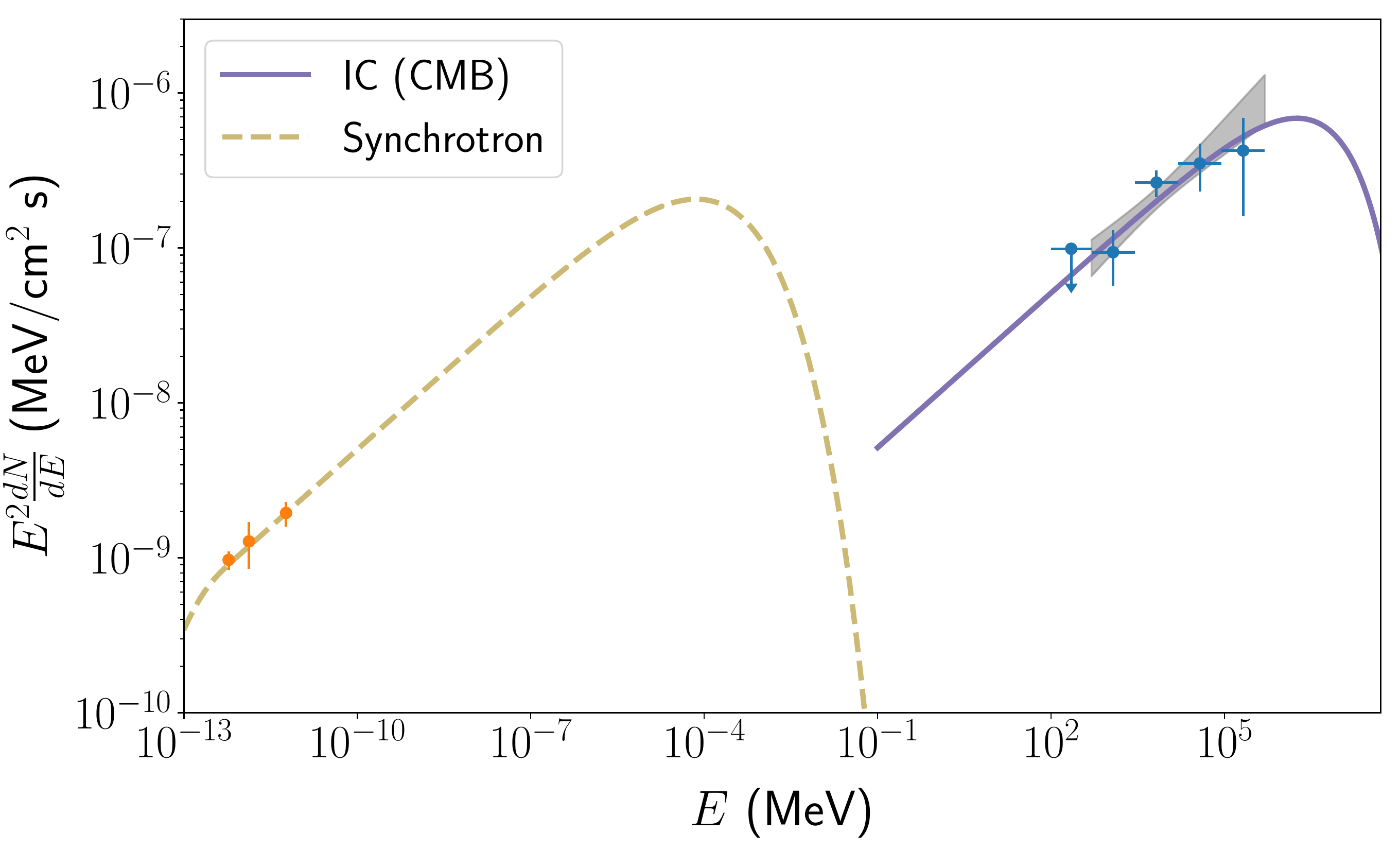}
    \caption{Non thermal fluxes measured from \source. The radio data were obtained from \protect\cite{2022arXiv220714141A}. The GeV fluxes from this work are shown with their $1\sigma$ statistical uncertainties. A 95\%-confidence level upper limit is shown for the energy interval 100--500~MeV. The dashed and solid lines represent the leptonic emission model applied to the data (see section \ref{discussion}). The gray band represents the $1\sigma$ propagated statistical uncertainty from the fit to the LAT data in the entire energy range used.}
    \label{fig:SED}
\end{figure}

We obtained the GeV spectral energy distribution (SED) fluxes of \source{} by dividing the LAT data in five energy bands and fitting the normalization in each bin. They are shown in Fig. \ref{fig:SED} with the radio continuum fluxes and the model applied in section \ref{discussion}. The LAT fluxes are also shown in Table \ref{tab2} as well as the TS value obtained for the disk template representing \source{} in each SED bin, and the 95\%-confidence level (CL) upper limit for the source flux in the energy interval 100--500~MeV.

\begin{table*}
\centering
\caption{Gamma-ray fluxes (and $1\sigma$ statistical uncertainties) and TS values for \source{} in each SED bin}
\begin{tabular}{|c|c|c|}
\hline
\hline
Energy range (GeV) & $E^2\frac{dN}{dE}$ ($10^{-7} $~MeV~cm$^{-2}$~s$^{-1}$) & TS\\
\hline
0.1--0.5  &  $0.99^a$ & $<4$\\
0.5--2.81  &  $0.94\pm 0.37$ & 7.6\\
2.81--15.8  &  $2.6\pm 0.5$ & 31.5\\
15.8--88.9  &  $3.5\pm 1.2$ & 19.1\\
88.9--500 &  $4.2\pm 2.6$ & 5.4\\
\hline
\end{tabular}\\
\textsuperscript{$a$}\footnotesize{95\%-CL upper limit on flux.}
\label{tab2}
\end{table*}

\section{Discussion}\label{discussion}
\subsection{On the origin of the $\gamma-$ray emission}
\cite{2022arXiv220714141A} present strong evidence that the radio emission they report is the remnant of the explosion that formed Calvera, but regardless of the origin of the Calvera pulsar we present in this work further evidence that the source is an SNR. The presence of high-energy emission in a region consistent with the size of the radio ring confirms the existence of relativistic particles in this object, as seen in other SNRs. Furthermore, the spectrum of the $\gamma-$ray emission is fully consistent with the measured radio spectrum, indicating that the same electrons producing the synchrotron emission at low energies, in the magnetic field of the source, are responsible for the GeV emission by inverse Compton (IC) scattering soft energy photons, particularly those of the Cosmic Microwave Background (CMB). From our measured GeV spectral index $\Gamma=1.66\pm 0.10_{\mbox{\tiny stat}} \pm 0.03_{\mbox{\tiny sys}}$ the predicted radio spectral index for the synchrotron emission ($S_\nu \propto \nu^{\alpha}$) from the same uncooled population of electrons under this model is $\alpha = 1-\Gamma \sim -0.7$, while the measured value is $-0.71\pm 0.09$. 

We used the {\tt naima} package \citep{naima} to fit the radio and GeV fluxes with a simple one-zone leptonic (IC-CMB) model. A particle distribution that is a power-law with an exponential cutoff is assumed, however the data are not able to constrain the cutoff energy. This is not surprising given that both non-thermal spectra in radio and $\gamma-$rays are described by simple power-law functions. Future TeV observations will be able to constrain this energy. Thus we fixed the electron cutoff energy at 50~TeV and fit the particle spectral normalization and index. These parameters depend on the value used for the cutoff energy and therefore should only be considered as representative values. The total energy content in the particles, on the other hand, for a given distance to the source, varies little with the cutoff energy chosen and thus can be used as a more reliable indication of a property of the source.

The results of the fit are shown in Fig. \ref{fig:SED}. The required magnetic field is $B=1.5^{+0.5}_{-0.3}\,\mu$G and the spectral index of the lepton distribution is $2.3^{+0.08}_{-0.07}$. The total energy content in the particles, integrated above a particle energy of 1\,GeV, is $(8.6^{+3.7}_{-2.6})\times 10^{47}\,\left( \frac{d}{3.3\, \mbox{\tiny kpc}} \right)^2$\,erg, where $d$ is the distance to the source. As noted earlier, a distance of 3.3~kpc was estimated recently for Calvera \citep{2021ApJ...922..253M}, and we used this as a reference value in the result. The total energy in the particles is very low: $\sim 0.1$\% of the typical kinetic energy available in SNRs for $d=3.3$~kpc. The typical available kinetic energy is more than enough to provide the necessary particle energy for any reasonable distance to the source. The $\gamma-$ray luminosity of \source{} is also relatively low. From the integrated GeV energy flux ($\int E\frac{dN}{dE}\,dE$) in the 1--500~GeV energy range we estimate a luminosity of $4.7\times10^{33}$~erg~s$^{-1}$, for a source distance of 3.3~kpc.

Another possible source of high-energy emission is a pulsar wind nebula (PWN). Powered by an energetic pulsar, the radiation from relativistic leptons in these systems is usually seen inside an SNR, and a neutron star (Calvera) is indeed likely associated to the SNR. In many systems the high-energy emission from a PWN is mostly due to IC emission and thus its $\gamma-$ray SED could be similar to that obtained here from LAT data. We have seen that the extent of the GeV emission is consistent with that of the radio emission (see Fig. \ref{fig:tsmap}) and that the radio spectrum correctly predicts the observed slope at GeV energies, thus a likely connection between the radio and $\gamma-$rays is established. In the present case, however, the morphology of the radio emission is shell-like, while the morphologies of the radio and high-energy emission from a PWN are usually centre filled. Regarding the possibility that the radio emission comes from the SNR shell while the GeV emission from a PWN, no composite SNRs are known for which the size of the PWN is as large as the radio extent of the host SNR. In addition, no evidence for non-thermal X-ray emission associated to Calvera, characteristic of a PWN, has been found \citep{2011MNRAS.410.2428Z}. It is therefore more likely that the high-energy electrons accelerated by the shock of the SNR are responsible for the observed $\gamma-$rays. The presence of these electrons and the CMB inevitably leads to the production of $\gamma-$rays.

\subsubsection{Hadronic scenario}
If we fit the GeV data with a hadronic scenario assuming the $\gamma-$rays result from inelastic collisions between relativistic protons and ambient protons in the form of gas (again using a power-law with an exponential cutoff and a fixed cutoff enery of 50~TeV for the particle distribution) the required spectral index for the particles is $1.80\pm0.10$. However, this scenario is problematic from the energetic point of view. \cite{2022arXiv220714141A} estimated an ambient gas density of $4\times 10^{-4}$~cm$^{-3}$ using the Galactic halo gas model of \cite{2013ApJ...770..118M}. If this value is adopted, the energetic requirement in the cosmic rays would be huge for a source distance of 3.3~kpc, of the order of $10^{53}$~erg, and basically impossible to be provided by a supernova explosion.

A much higher ambient gas density than predicted by standard Galactic halo gas models is needed in a hadronic scenario. For an average gas number density of 0.1~cm$^{-3}$ , the required total energy in the particles is $\sim2\times 10^{50}$~erg, which is closer to the standard 10\% of the typical kinetic energy content in the SNR shock expected to be transferred to the cosmic rays. More observations are necessary to probe the environment of the source in order to compare the actual density with the prediction from the hadronic model.

In addition, in the hadronic scenario, a harder proton distribution than predicted by linear diffusive shock acceleration would be required to explain the GeV emission from \source{}, while hadronic emission detected from SNRs typically shows a softer spectrum (see below). For this reason we believe that the hadronic scenario is less likely to be responsible for the $\gamma-$rays.

\subsection{Comparison to other SNRs}
It is interesting to compare the known properties of this source with those of G17.8$+$16.7, a recently confirmed SNR also located outside the Galactic plane \citep{2022MNRAS.510.2920A}. This source shows a similar radio extension as well as similar (although slightly softer) radio and GeV spectra. Its high-energy luminosity is also similar to that of \source. The radio and GeV SED of G17.8$+$16.7 can also be naturally explained by leptonic emission, which at $\gamma-$ray energies is expected to dominate over any hadronic contribution due to the low density environment in which the SNR is presumably expanding outside the Galactic plane. As pointed out by these authors, other radio-dim SNRs show similar GeV spectra such as G279.0$+$1.1 \citep{2020MNRAS.492.5980A}, G150.3$+$4.5 \citep{2014A&A...567A..59G,2020A&A...643A..28D} and G$323.7-1.0$ \citep{2017ApJ...843...12A}, while other unknown sources detected only at GeV or TeV energies with similar GeV features, such as G350.6$-$4.7 and 2HWC~J2006$+$341, could also turn out to be dim SNRs \citep{2018MNRAS.474..102A,2020ApJ...903L..14A}.

Some simulations show that kyr-old SNRs evolving in low-density circumstelar environments could have very low synchrotron fluxes (compared to most known SNRs) while their SEDs are dominated by IC emission \citep{2019ApJ...876...27Y}, and the particles would also experience low synchrotron losses and live longer. This could also apply to SNRs expanding in very low density environments, which is very likely the case for \source{} and G17.8$+$16.7. In contrast, GeV-emitting SNRs that are expanding into dense gas have $\gamma-$ray luminosities that are one to two orders of magnitude higher than those of \source{} and G17.8$+$16.7, and a spectrum characteristic of hadronic emission \citep[e.g.,][]{2013Sci...339..807A}. Further observations are important to constrain the parameters of both of these SNRs and of their environments to understand their evolution.

If located at a distance of 3.3~kpc, the radius of the SNR would be about 27~pc. Applying the Sedov-Taylor self-similar solution, adopting the ambient density estimated by \cite{2022arXiv220714141A} and an explosion energy of $10^{51}$~erg, we find an age of $\sim1.4$~kyr for the remnant (note that these authors estimate an age of 7.7~kyr using the same assumptions but an SNR radius of 54~pc). For a relatively young age of 1.4~kyr a historical record of the supernova event would be more likely. However, this age estimate can after all be a bad estimate of the true age if the self-similar solution does not apply, which is possible for SNRs expanding in low-density media such as in the Galactic halo \citep{2005ApJ...628..205T}. The synchrotron cooling time of TeV electrons in a magnetic field of $1.5\,\mu$G is of the order of $\sim$~Myr while the time it takes for an old SNR to dissipate is much smaller, so particles could very well survive for as long as they are confined in the remnant.

For a distance of 3.3~kpc to the source, its 1.4~GHz luminosity would be $\sim3\times10^{14}$~W~Hz$^{-1}$ which is comparable to the lowest luminosities seen in SNRs in the Local Group and Magellanic Clouds \citep[e.g.,][]{1998ApJ...504..761C}. Deeper surveys are needed to reveal previously unknown low-surface brightness SNRs. As seen in this and other studies, $\gamma-$ray surveys could also contribute to probe this population and help understand their evolution and properties.

\section{Summary}
We have confirmed the existence of relativistic particles at the location of the non-thermal radio ring recently discovered around the Calvera pulsar at high Galactic latitude, confirming the nature of the source as an SNR, \source. The morphology of the $\gamma-$ray emission produced by these particles is consistent with the extent of the radio emission and its spectrum is naturally explained by leptonic emission (inverse Compton scattering of CMB photons by the same electrons emitting the synchrotron radiation). A hadronic origin for the GeV emission requires an enhancement of the ambient density by several orders of magnitude compared to standard models of the Galactic halo gas as well as a harder particle distribution than predicted by linear diffusive shock acceleration theory. An origin in a PWN is disfavored mainly from the source morphology. The similar radio and GeV properties of \source{} and other radio-dim SNRs, particularly G17.8$+$16.7, recently found outside the Galactic plane, reveal the importance deep radio and $\gamma-$ray surveys have in understanding the properties of the dim SNR population.
 
\section*{Acknowledgements}
We thank the anonymous referee for useful comments that helped improve this work, as well as funding from Universidad de Costa Rica under grant B8267. This research has made use of NASA's Astrophysics Data System.

\section*{Data availability.} This work makes use of publicly available \textit{Fermi}-LAT data provided online by the Fermi Science Support Center at \\ http://fermi.gsfc.nasa.gov/ssc/.

\bibliographystyle{mnras}
\bibliography{calveraSNR}

\begin{thebibliography}{}
\makeatletter
\relax
\def\mn@urlcharsother{\let\do\@makeother \do\$\do\&\do\#\do\^\do\_\do\%\do\~}
\def\mn@doi{\begingroup\mn@urlcharsother \@ifnextchar [ {\mn@doi@}
  {\mn@doi@[]}}
\def\mn@doi@[#1]#2{\def\@tempa{#1}\ifx\@tempa\@empty \href
  {http://dx.doi.org/#2} {doi:#2}\else \href {http://dx.doi.org/#2} {#1}\fi
  \endgroup}
\def\mn@eprint#1#2{\mn@eprint@#1:#2::\@nil}
\def\mn@eprint@arXiv#1{\href {http://arxiv.org/abs/#1} {{\tt arXiv:#1}}}
\def\mn@eprint@dblp#1{\href {http://dblp.uni-trier.de/rec/bibtex/#1.xml}
  {dblp:#1}}
\def\mn@eprint@#1:#2:#3:#4\@nil{\def\@tempa {#1}\def\@tempb {#2}\def\@tempc
  {#3}\ifx \@tempc \@empty \let \@tempc \@tempb \let \@tempb \@tempa \fi \ifx
  \@tempb \@empty \def\@tempb {arXiv}\fi \@ifundefined
  {mn@eprint@\@tempb}{\@tempb:\@tempc}{\expandafter \expandafter \csname
  mn@eprint@\@tempb\endcsname \expandafter{\@tempc}}}

\bibitem[\protect\citeauthoryear{{Abdollahi} et~al.,}{{Abdollahi}
  et~al.}{2020}]{2020ApJS..247...33A}
{Abdollahi} S.,  et~al., 2020, \mn@doi [\apjs] {10.3847/1538-4365/ab6bcb},
  \href {https://ui.adsabs.harvard.edu/abs/2020ApJS..247...33A} {247, 33}

\bibitem[\protect\citeauthoryear{{Abdollahi} et~al.,}{{Abdollahi}
  et~al.}{2022}]{2022ApJS..260...53A}
{Abdollahi} S.,  et~al., 2022, \mn@doi [\apjs] {10.3847/1538-4365/ac6751},
  \href {https://ui.adsabs.harvard.edu/abs/2022ApJS..260...53A} {260, 53}

\bibitem[\protect\citeauthoryear{{Acero} et~al.,}{{Acero}
  et~al.}{2016}]{2016ApJS..224....8A}
{Acero} F.,  et~al., 2016, \mn@doi [\apjs] {10.3847/0067-0049/224/1/8}, \href
  {https://ui.adsabs.harvard.edu/abs/2016ApJS..224....8A} {224, 8}

\bibitem[\protect\citeauthoryear{{Ackermann} et~al.,}{{Ackermann}
  et~al.}{2012}]{2012ApJS..203....4A}
{Ackermann} M.,  et~al., 2012, \mn@doi [\apjs] {10.1088/0067-0049/203/1/4},
  \href {https://ui.adsabs.harvard.edu/abs/2012ApJS..203....4A} {203, 4}

\bibitem[\protect\citeauthoryear{{Ackermann} et~al.,}{{Ackermann}
  et~al.}{2013}]{2013Sci...339..807A}
{Ackermann} M.,  et~al., 2013, \mn@doi [Science] {10.1126/science.1231160},
  \href {https://ui.adsabs.harvard.edu/abs/2013Sci...339..807A} {339, 807}

\bibitem[\protect\citeauthoryear{{Akaike}}{{Akaike}}{1974}]{1974ITAC...19..716A}
{Akaike} H.,  1974, IEEE Transactions on Automatic Control, \href
  {http://adsabs.harvard.edu/abs/1974ITAC...19..716A} {19, 716}

\bibitem[\protect\citeauthoryear{{Albert} et~al.,}{{Albert}
  et~al.}{2020}]{2020ApJ...903L..14A}
{Albert} A.,  et~al., 2020, \mn@doi [\apjl] {10.3847/2041-8213/abbfae}, \href
  {https://ui.adsabs.harvard.edu/abs/2020ApJ...903L..14A} {903, L14}

\bibitem[\protect\citeauthoryear{{Anderson} et~al.,}{{Anderson}
  et~al.}{2017}]{2017A&A...605A..58A}
{Anderson} L.~D.,  et~al., 2017, \mn@doi [\aap] {10.1051/0004-6361/201731019},
  \href {https://ui.adsabs.harvard.edu/abs/2017A&A...605A..58A} {605, A58}

\bibitem[\protect\citeauthoryear{{Araya}}{{Araya}}{2017}]{2017ApJ...843...12A}
{Araya} M.,  2017, \mn@doi [\apj] {10.3847/1538-4357/aa7261}, \href
  {https://ui.adsabs.harvard.edu/abs/2017ApJ...843...12A} {843, 12}

\bibitem[\protect\citeauthoryear{{Araya}}{{Araya}}{2018}]{2018MNRAS.474..102A}
{Araya} M.,  2018, \mn@doi [\mnras] {10.1093/mnras/stx2779}, \href
  {https://ui.adsabs.harvard.edu/abs/2018MNRAS.474..102A} {474, 102}

\bibitem[\protect\citeauthoryear{{Araya}}{{Araya}}{2020}]{2020MNRAS.492.5980A}
{Araya} M.,  2020, \mn@doi [\mnras] {10.1093/mnras/staa244}, \href
  {https://ui.adsabs.harvard.edu/abs/2020MNRAS.492.5980A} {492, 5980}

\bibitem[\protect\citeauthoryear{{Araya}, {Hurley-Walker}  \&
  {Quir{\'o}s-Araya}}{{Araya} et~al.}{2022}]{2022MNRAS.510.2920A}
{Araya} M.,  {Hurley-Walker} N.,   {Quir{\'o}s-Araya} S.,  2022, \mn@doi
  [\mnras] {10.1093/mnras/stab3550}, \href
  {https://ui.adsabs.harvard.edu/abs/2022MNRAS.510.2920A} {510, 2920}

\bibitem[\protect\citeauthoryear{{Arias} et~al.,}{{Arias}
  et~al.}{2022}]{2022arXiv220714141A}
{Arias} M.,  et~al., 2022, arXiv e-prints, \href
  {https://ui.adsabs.harvard.edu/abs/2022arXiv220714141A} {p. arXiv:2207.14141}

\bibitem[\protect\citeauthoryear{{Atwood} et~al.,}{{Atwood}
  et~al.}{2009}]{2009ApJ...697.1071A}
{Atwood} W.~B.,  et~al., 2009, \mn@doi [\apj] {10.1088/0004-637X/697/2/1071},
  \href {https://ui.adsabs.harvard.edu/abs/2009ApJ...697.1071A} {697, 1071}

\bibitem[\protect\citeauthoryear{{Bogdanov} et~al.,}{{Bogdanov}
  et~al.}{2019}]{2019ApJ...877...69B}
{Bogdanov} S.,  et~al., 2019, \mn@doi [\apj] {10.3847/1538-4357/ab1b2e}, \href
  {https://ui.adsabs.harvard.edu/abs/2019ApJ...877...69B} {877, 69}

\bibitem[\protect\citeauthoryear{{Burnham} \& {Anderson}}{{Burnham} \&
  {Anderson}}{2001}]{burnham2001}
{Burnham} K.~P.,  {Anderson} D.~R.,  2001, \mn@doi [Wildlife Research]
  {10.1071/WR99107}, 28, 111

\bibitem[\protect\citeauthoryear{{Case} \& {Bhattacharya}}{{Case} \&
  {Bhattacharya}}{1998}]{1998ApJ...504..761C}
{Case} G.~L.,  {Bhattacharya} D.,  1998, \mn@doi [\apj] {10.1086/306089}, \href
  {https://ui.adsabs.harvard.edu/abs/1998ApJ...504..761C} {504, 761}

\bibitem[\protect\citeauthoryear{{Devin}, {Lemoine-Goumard}, {Grondin},
  {Castro}, {Ballet}, {Cohen}  \& {Hewitt}}{{Devin}
  et~al.}{2020}]{2020A&A...643A..28D}
{Devin} J.,  {Lemoine-Goumard} M.,  {Grondin} M.~H.,  {Castro} D.,  {Ballet}
  J.,  {Cohen} J.,   {Hewitt} J.~W.,  2020, \mn@doi [\aap]
  {10.1051/0004-6361/202038503}, \href
  {https://ui.adsabs.harvard.edu/abs/2020A&A...643A..28D} {643, A28}

\bibitem[\protect\citeauthoryear{{Gao} \& {Han}}{{Gao} \&
  {Han}}{2014}]{2014A&A...567A..59G}
{Gao} X.~Y.,  {Han} J.~L.,  2014, \mn@doi [\aap] {10.1051/0004-6361/201424128},
  \href {https://ui.adsabs.harvard.edu/abs/2014A&A...567A..59G} {567, A59}

\bibitem[\protect\citeauthoryear{{Green}}{{Green}}{2019}]{2019JApA...40...36G}
{Green} D.~A.,  2019, \mn@doi [Journal of Astrophysics and Astronomy]
  {10.1007/s12036-019-9601-6}, \href
  {https://ui.adsabs.harvard.edu/abs/2019JApA...40...36G} {40, 36}

\bibitem[\protect\citeauthoryear{{Green}, {Reeves}  \& {Murphy}}{{Green}
  et~al.}{2014}]{2014PASA...31...42G}
{Green} A.~J.,  {Reeves} S.~N.,   {Murphy} T.,  2014, \mn@doi [\pasa]
  {10.1017/pasa.2014.37}, \href
  {https://ui.adsabs.harvard.edu/abs/2014PASA...31...42G} {31, e042}

\bibitem[\protect\citeauthoryear{{Haberl}}{{Haberl}}{2007}]{2007Ap&SS.308..181H}
{Haberl} F.,  2007, \mn@doi [\apss] {10.1007/s10509-007-9342-x}, \href
  {https://ui.adsabs.harvard.edu/abs/2007Ap&SS.308..181H} {308, 181}

\bibitem[\protect\citeauthoryear{{Halpern}, {Bogdanov}  \&
  {Gotthelf}}{{Halpern} et~al.}{2013}]{2013ApJ...778..120H}
{Halpern} J.~P.,  {Bogdanov} S.,   {Gotthelf} E.~V.,  2013, \mn@doi [\apj]
  {10.1088/0004-637X/778/2/120}, \href
  {https://ui.adsabs.harvard.edu/abs/2013ApJ...778..120H} {778, 120}

\bibitem[\protect\citeauthoryear{{Hurley-Walker} et~al.,}{{Hurley-Walker}
  et~al.}{2019}]{2019PASA...36...48H}
{Hurley-Walker} N.,  et~al., 2019, \mn@doi [\pasa] {10.1017/pasa.2019.33},
  \href {https://ui.adsabs.harvard.edu/abs/2019PASA...36...48H} {36, e048}

\bibitem[\protect\citeauthoryear{{Lande} et~al.,}{{Lande}
  et~al.}{2012}]{2012ApJ...756....5L}
{Lande} J.,  et~al., 2012, \mn@doi [\apj] {10.1088/0004-637X/756/1/5}, \href
  {https://ui.adsabs.harvard.edu/abs/2012ApJ...756....5L} {756, 5}

\bibitem[\protect\citeauthoryear{{Li}, {Wheeler}, {Bash}  \& {Jefferys}}{{Li}
  et~al.}{1991}]{1991ApJ...378...93L}
{Li} Z.,  {Wheeler} J.~C.,  {Bash} F.~N.,   {Jefferys} W.~H.,  1991, \mn@doi
  [\apj] {10.1086/170409}, \href
  {https://ui.adsabs.harvard.edu/abs/1991ApJ...378...93L} {378, 93}

\bibitem[\protect\citeauthoryear{{Mattox} et~al.,}{{Mattox}
  et~al.}{1996}]{1996ApJ...461..396M}
{Mattox} J.~R.,  et~al., 1996, \mn@doi [\apj] {10.1086/177068}, \href
  {http://adsabs.harvard.edu/abs/1996ApJ...461..396M} {461, 396}

\bibitem[\protect\citeauthoryear{{Mereghetti}, {Rigoselli}, {Taverna},
  {Baldeschi}, {Crestan}, {Turolla}  \& {Zane}}{{Mereghetti}
  et~al.}{2021}]{2021ApJ...922..253M}
{Mereghetti} S.,  {Rigoselli} M.,  {Taverna} R.,  {Baldeschi} L.,  {Crestan}
  S.,  {Turolla} R.,   {Zane} S.,  2021, \mn@doi [\apj]
  {10.3847/1538-4357/ac34f2}, \href
  {https://ui.adsabs.harvard.edu/abs/2021ApJ...922..253M} {922, 253}

\bibitem[\protect\citeauthoryear{{Miller} \& {Bregman}}{{Miller} \&
  {Bregman}}{2013}]{2013ApJ...770..118M}
{Miller} M.~J.,  {Bregman} J.~N.,  2013, \mn@doi [\apj]
  {10.1088/0004-637X/770/2/118}, \href
  {https://ui.adsabs.harvard.edu/abs/2013ApJ...770..118M} {770, 118}

\bibitem[\protect\citeauthoryear{{Shimwell} et~al.,}{{Shimwell}
  et~al.}{2017}]{2017A&A...598A.104S}
{Shimwell} T.~W.,  et~al., 2017, \mn@doi [\aap] {10.1051/0004-6361/201629313},
  \href {https://ui.adsabs.harvard.edu/abs/2017A&A...598A.104S} {598, A104}

\bibitem[\protect\citeauthoryear{{Tammann}, {Loeffler}  \&
  {Schroeder}}{{Tammann} et~al.}{1994}]{1994ApJS...92..487T}
{Tammann} G.~A.,  {Loeffler} W.,   {Schroeder} A.,  1994, \mn@doi [\apjs]
  {10.1086/192002}, \href
  {https://ui.adsabs.harvard.edu/abs/1994ApJS...92..487T} {92, 487}

\bibitem[\protect\citeauthoryear{{Tang} \& {Wang}}{{Tang} \&
  {Wang}}{2005}]{2005ApJ...628..205T}
{Tang} S.,  {Wang} Q.~D.,  2005, \mn@doi [\apj] {10.1086/430875}, \href
  {https://ui.adsabs.harvard.edu/abs/2005ApJ...628..205T} {628, 205}

\bibitem[\protect\citeauthoryear{{Wilks}}{{Wilks}}{1938}]{wilks1938}
{Wilks} S.~S.,  1938, \mn@doi [Ann. Math. Statist.] {10.1214/aoms/1177732360},
  9, 60

\bibitem[\protect\citeauthoryear{{Yasuda} \& {Lee}}{{Yasuda} \&
  {Lee}}{2019}]{2019ApJ...876...27Y}
{Yasuda} H.,  {Lee} S.-H.,  2019, \mn@doi [\apj] {10.3847/1538-4357/ab13ab},
  \href {https://ui.adsabs.harvard.edu/abs/2019ApJ...876...27Y} {876, 27}

\bibitem[\protect\citeauthoryear{{Zabalza}}{{Zabalza}}{2015}]{naima}
{Zabalza} V.,  2015, Proc.~of International Cosmic Ray Conference 2015, \href
  {http://adsabs.harvard.edu/abs/2015arXiv150903319Z} {p.~922}

\bibitem[\protect\citeauthoryear{{Zane} et~al.,}{{Zane}
  et~al.}{2011}]{2011MNRAS.410.2428Z}
{Zane} S.,  et~al., 2011, \mn@doi [\mnras] {10.1111/j.1365-2966.2010.17619.x},
  \href {https://ui.adsabs.harvard.edu/abs/2011MNRAS.410.2428Z} {410, 2428}

\makeatother
\end{thebibliography}

% Don't change these lines
\bsp    % typesetting comment
\label{lastpage}

\end{document}